\RequirePackage{fix-cm}
\documentclass{svjour3}                     % onecolumn (standard format)
\smartqed  % flush right qed marks, e.g. at end of proof
\usepackage{graphicx}
\usepackage{fullpage}
\usepackage{amsmath}
\usepackage{fancyvrb}
\usepackage{moreverb}

\DeclareMathOperator*{\argmin}{arg\,min} % thin space, limits underneath in displays

\usepackage{listings}        

\begin{document}

\title{Enhancing Keyword Correlation for Event Detection in Social Networks using SVD and K-Means: Twitter Case Study}%\thanks{Grants or other notes

\author{Ahmad Hany Hossny         \and
        Terry Moschou  \and
        Grant Osborne  \and
        Lewis Mitchell  \and
        Nick Lothian  \and
}

\institute{A. Hossny
\at
              School of Mathematical Sciences, Univeristy of Adelaide, Australia \\
              \email{ahmad.hossny@adelaide.edu.au}   
           \and
           T. Moschou, G. Osborne, N. Lothian \at
              Data to Decisions Cooperative Research Centre (D2D CRC), Adelaide, Australia \\
              \email{terry.moschou@d2dcrc.com.au, nick.lothian@d2dcrc.com.au, grant.osborne@d2dcrc.com.au} 
           \and
              L. Mitchell 
            \at
                           School of Mathematical Sciences, Univeristy of Adelaide, Australia\\
               D2D CRC Stream Lead\\
 \email{lewis.mitchell@adelaide.edu.au}
              }

\date{Received: date / Accepted: date}

\maketitle

\begin{abstract}
Extracting textual features from tweets is a challenging process due to the noisy nature of the content and the weak signal of most of the words used. In this paper, we propose using singular value decomposition (SVD) with clustering to enhance the signals of the textual features in the tweets to improve the correlation with events. The proposed technique applies SVD to the time series vector for each feature to factorize the matrix of feature/day counts, in order to ensure the independence of the feature vectors. Afterwards, the K-means clustering is applied to build a look-up table that maps members of each cluster to the cluster-centroid. The lookup table is used to map each feature in the original data to the centroid of its cluster, then we calculate the sum of the term frequency vectors of all features in each cluster to the term-frequency-vector of the cluster centroid. To test the technique we calculated the correlations of the cluster centroids with the golden standard record (GSR) vector before and after summing the vectors of the cluster members to the centroid-vector. The proposed method is applied to multiple correlation techniques including the Pearson, Spearman, distance correlation and Kendal Tao. The experiments have also considered the different word forms and lengths of the features including keywords, n-grams, skip-grams and bags-of-words. The correlation results are enhanced significantly as the highest correlation scores have increased from 0.3 to 0.6, and the average correlation scores have increased from 0.3 to 0.4.

\end{abstract}

\section{Introduction}
\label{intro}
Social networks such as Twitter and Facebook are frequently used to organize protests, rallies, and revolutions. Social events such as protests can be organized through the follower-followee scheme or through the spontaneous propagation scheme \cite{GonzalezBailon201695,Lee2015879}. The follower-followee scheme has a leader that calls his followers to a specific protest at a specific place and time. This pattern is easy to detect by tracking the effective leaders, those with a large number of followers, assuming their identities are known in advance. On the other hand, the spontaneous propagation scheme is initiated by multiple standard users with limited followers and impact, who speak out for their cause. The initial messages are propagated through close friends and followers to spread the word at an exponentially growing rate \cite{tufekci2012social,anduiza2014mobilization,valenzuela2013unpacking}. Identifying these events requires tracking the growth in usage rate for one or more keywords that are sufficiently associated with protests and rallies.

%Keywords challenges
Using Twitter text as features is challenging for multiple reasons such as the limited length of each tweet, the informal nature of the tweets and the multilingual nature of Twitter \cite{Fung:2005:PFB:1083592.1083616,Mathioudakis:2010:TTD:1807167.1807306,Petrovic:2010:SFS:1857999.1858020}. Tackling twitter challenges can be performed via NLP preprocessing steps such as lemmatization, stemming, lexical analysis, morphological analysis, syntactic analysis and Parts-Of-Speech tagging\cite{hossny2008automatic}. These tasks can be performed using rule-based techniques or machine learning techniques such as Inductive logic programming or deep neural networks depending on the amount of data to be processed\cite{hossny2009machine,azzam2017question}. The main challenges affecting text mining in twitter are listed below:
\begin{itemize}
    \item The tweet length of 140 characters makes topic modelling and sentiment analysis very challenging for individual tweets. 
    \item The frequent usage of acronyms, misspelled words and non-standard abbreviations make many words difficult to detect.
    \item Using Roman script to write non-English language distorts the feature signals due to similar words from other languages (e.g. the term ``boss'' means ``look'' in Arabic, while in English it means ``manager'')
    \item Semantic ambiguity: Many words have multiple meanings (e.g. ``strike'' may refer to a protest, a lightning strike or a football strike).
    \item Synonyms: Conversely, the same meaning can be expressed by multiple words (e.g. the terms ``protest'' and ``rally'' are used interchangeably).
\end{itemize}

The correlation between the social events vector and the time series of the keyword frequency is affected by three factors. The first factor is the word-form, whether single words, n-grams, skip-grams or bags-of-words (BOWs) \cite{fernandez2014gplsi}. The second factor is the number of words used as a feature in the n-gram, skip-gram and the BOWs \cite{li2012twevent,martinez2015ensemble}. The third factor is the correlation technique used, such as Pearson, Spearman, distance-correlation, or mutual information \cite{Eysenbach2011,Riquelme2016949}. The combination of word-forms, word-counts and correlation techniques selects different sets of words as the best features to identify civil unrest events. In this paper, we apply the proposed technique in experiments involving all of the mentioned word forms, word counts and correlation techniques. Correlation scores were improved for most of the experiments with different ratios.

\begin{figure*}
\centering
  \includegraphics[width=0.70\textwidth]{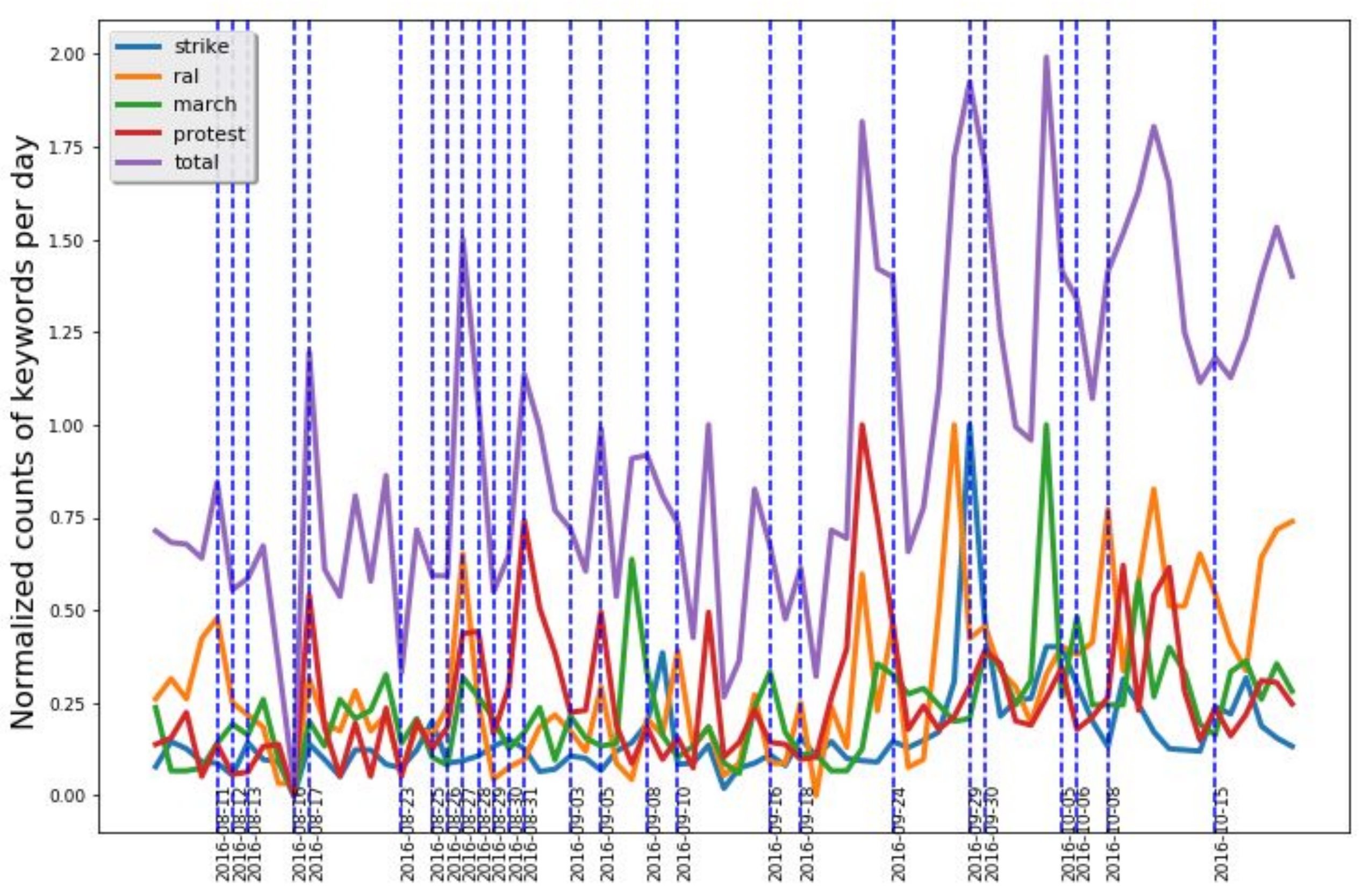}
% Figure caption is below the Figure
\caption{The count vector for each word of `protest', `rally', `strike' and `march' shows few spikes matching the days of the events (vertical dashed lines). The summed signal of the four words can match the event days with spikes more than any of the individual words, and also has fewer mismatching spikes than any of the individual words}
\label{fig:wordsignals}       % Give a unique label
\end{figure*}

%LSA, topic modelling, KSVD

This research aims to improve the correlation between textual features and events by calculating the sum of the time series vectors of multiple features having similar meanings, to form a single feature that represents all of its constituent time series vectors. This representative feature is selected by clustering the features and finding the one with the minimum distance to all others, which is known as the cluster centroid. So, we transform the features using SVD and cluster the transformed features to build the lookup table mapping the features labels (e.g. Bags-of-Words) to the centroid feature label. Then we use the lookup table to know which vectors of time series raw counts (not-transformed) should be summed up.  Adding the related features to the centroid feature improves the correlation score for the centroid feature significantly without affecting the correlation score of the other features within the cluster.

Enhancing the correlation will give us more informative features with stronger signals that will allow us to perform live detection of ongoing events as soon as it occurs. In this paper, we use civil unrest events in Melbourne as a case study for correlating keywords with social events, then use these keywords to detect protest immideatley as soon it occur, once the keywords of interest occur more than a specific threshold. Here, we consider the golden standard records as a count vector describing how many civil unrest events happened in a specific day within the time-frame, which manually curated from news article along the timeframe of the experiment. We also consider the features as the vectors describing the daily counts of each of keywords, n-grams, skip-grams or bags-of-words(BOWs) along the time-frame. In the proposed method, we aim to combine the vectors of related BOWs having similar meaning in the context of event (civil unrest) such as the BOWs of (``Melbourne'' , ``protest''), (``Melbourne'',``rally''), (``Melbourne'',``strike'') and (``Melbourne'',``march''), as each of them has a relatively weak signal, which means a small magnitude for the daily counts within each vector . Then, Combine the signals of the four BOWs into one BOw gives a stronger signal that has the same meaning and higher matching scores with events, as indicated in Figure \ref{fig:wordsignals}.  

The proposed technique is to use singular value decomposition (SVD) to factorize the feature/day matrix into a feature matrix, the daily event matrix and the singular matrix mapping the features to the events  \cite{golub1970singular,klema1980singular,lange2010singular}. SVD is important to ensure that features' locations in the space are mapped to orthogonal dimensions, as K-means uses Euclidean distance and requires an orthogonal relation among the features. This orthogonality is not guaranteed in the original matrix, as the textual features are not guaranteed to be independent of each other and neither are the days. Once the SVD is applied, the resulting matrices are guaranteed to represent the features (BOWs) as orthogonal vectors in the features matrix and the days are represented as orthogonal vectors in the observation matrix. After decomposition, the feature matrix is clustered using k-means and the centroids of the clusters are used as the master feature for correlation with the event vector. 

%Paper layout
In section 2 we describe the most recognized feature extraction techniques. Section 3 explains the proposed technique including the SVD and how it is applied to our problem. Section 4 will explain the experiments and the results. Section 5 will state our conclusion and directions for future work.
\section{Feature Extraction Techniques}
\label{sec:1}
Feature extraction is the process of preparing the features selected from data to be used for training the learning model. Feature extraction aims to reduce computational complexity, eliminate misleading features and strengthen weak signals. Computational complexity is reduced through dimensionality reduction. Misleading features can be eliminated through filtration according to the frequency range, the variance or the signal-to-noise-ratio. Weak signals are improved by combining multiple features into one via clustering. The feature extraction process can be performed in geometric space using PCA or SVD by transforming the feature vectors into orthogonal vectors that can be projected, eliminated or clustered as needed. Latent Semantic Indexing (LSI) is an example of feature reduction via projection \cite{evangelopoulos2013latent}, and K-SVD is an example of improving the signal via clustering \cite{jiang2013label}. Many techniques have been proposed for feature reduction including Principal Component Analysis (PCA), Singular-Value-Decomposition (SVD), Independent Component Analysis (ICA), Common-Spatial-Patterns (CSP), and Latent Dirichlet Allocation (LDA). We will describe these briefly in the following subsections.

\subsection{Principal Component Analysis}
\label{sec:PCA}
Principal Component Analysis (PCA) is the process of finding the best linear subspace, where the first component is a straight line with smallest orthogonal distance to all points. PCA ranks the features according to their variance in descending order, where the new components are orthogonal to each other. PCA is performed using eigenvalue decomposition of the covariance matrix for the feature/observation matrix. This process results in two matrices, the first is the set of eigenvectors and the other matrix is diagonal with eigenvalues $\lambda_i$ in decreasing order along the diagonal \cite{wold1987principal,abdi2010principal}. 

The goal of PCA is to ensure that each of the feature vectors (eigenvectors in the first matrix) is independent and orthogonal to the other features. This makes the process of projecting higher dimensions onto lower dimensions applicable. Meanwhile, sorting eigenvalues in the second matrix in a descending order simplifies the feature reduction process, as the smallest eigenvalues indicate the least significant features, which can be eliminated. PCA is also described as rotation, scaling, and projection of the original matrix to match the reduced matrix where all the vectors are orthogonal \cite{shlens2014tutorial}. 

 The applicability of PCA is limited by the assumption of linearity, as it simplifies the problem by limiting the basis and by formalizing the continuity assumption. This assumption limits PCA to representing the data as a linear combination of its features \cite{Spiegelberg201740}. PCA has been used frequently to enhance signals or to increase the signal to noise ratio in fields such as image processing \cite{potapov2017enhancement}, medical imaging (fMRI and XRAY) \cite{Soltysik201518,chen2005enhancing}, control theory \cite{hamadache2017principal}, remote sensing \cite{koutsias2009forward} and neuro-computing \cite{yu2014analysis,sun2008random}.

\subsection{Singular Value Decomposition}
\label{sec:SVD}
Singular value decomposition is the process of factorizing the feature/document matrix into three matrices. The first matrix represents the features, the third matrix represents the documents and the matrix in between is a diagonal matrix that maps the features to the documents \cite{EWERBRING198937,golub1970singular}. The two matrices resulting from the SVD consist of orthonormal vectors, which makes distance measurement between vectors in the same matrix possible using Euclidean distance or cosine similarity. This concept is applied in Latent Semantic Indexing (LSI) that is used in recommender systems, and we apply the same concept to clustering as well. SVD is considered an extended version of PCA, as the feature matrix resulting from SVD is exactly the same eigenvector matrix that results from PCA, enabling SVD to be used for feature reduction similarly to PCA \cite{Wall2003}.

\subsection{Latent Semantic Indexing}
\label{sec:LSA}
Latent Semantic Indexing (LSI) or Latent Semantic Analysis (LSA) is a method to analyze the relationships between documents and their word contents using a set of mapping concepts. LSA assumes that text follows the distributional hypothesis, where words with similar meanings will appear in similar contexts with similar distribution \cite{landauer2006latent}. So, LSA formulates the term frequency per document as a matrix with rows representing words and columns representing documents. LSI uses SVD to decompose the term-frequency matrix into the orthonormal term matrix, the orthonormal document matrix and the concept-mapping matrix. LSA can be used to reduce the number of terms used as features in the first matrix \cite{dumais2004latent}, or to evaluate two documents' similarity by calculating the cosine similarity of any two vectors in the document matrix.

%\subsection{Other techniques}
\subsection{Indepenedent Component Analysis}
\label{sec:ICA}
Independent Component Analysis (ICA)  is a statistical technique that utilizes a mix of PCA and factor analysis to find the latent variables controlling a set of observations. This technique assumes the observations are linear mixtures of non-Gaussian and mutually independent latent variables \cite{hyvarinen2004independent}, and finds statistically independent features regardless of their influence on the response variable \cite{hyvarinen2000independent}. ICA transforms the feature space linearly into a new feature space, where each of the new features is statistically independent of any other transformed features. This transformation makes the mutual information of any two vectors equal to zero and the mutual information of the two-feature matrix as high as possible \cite{comon1994independent}.

\subsection{Common Spatial Pattern}
\label{sec:CSP}
Common spatial pattern (CSP) is a feature extraction technique that learns spatial filters from the data by maximizing the variance of filtered signals in the first class and minimizing the variance of the other class  \cite{ramoser2000optimal,blankertz2008optimizing}. CSP is similar to ICA as it decomposes the multivariate signal into multiple additive sub-signals with maximum differences in variance between two classes  \cite{lotte2011regularizing}. CSP is usually used in binary classification, and it can be extended for multiple classifications by following the one-vs-rest scheme. CSP is sensitive to noise and can overfit easily with small sets of training data. The objective of CSP  is to achieve the optimal classification for the signal using the band power features \cite{diggle2013statistical}.

%\subsection{Canonical Correlation Analysis}
%\label{sec:CCA}

\subsection{Latent Dirichlet allocation  }
\label{sec:LDA}
Latent Dirichlet Allocation (LDA) is a generative probabilistic model that is used frequently in topic modelling. It represents the documents as a random mix of latent topics \cite{blei2003latent}, with each topic identified by the distribution of the used words. LDA is formulated as a Bayesian model of three levels, where each document is modelled as a mix of underlying topics and each topic is also modelled as a mix of underlying probabilities of words \cite{hoffman2010online}. LDA is used frequently for feature extraction, as \cite{wang2012automatic} used it to reduce the features for crime prediction using twitter. It has also been used for tracking user interests in Twitter by \cite{sasaki2014online}.

%The LDA model is a Bayesian version of Latent Semantic Indexing (LSI) with better performance on a smaller amount of %of data, especially that Bayesian techniques are immune to overfitting \cite{niraula2013experiments}. The two %models can achieve similar results for the large datasets. LSI has the limitation that adding a new document will %require retraining the re-fit whole model \cite{biro2008comparative}.

\section{The Proposed Technique: Decompose-Cluster-Map}
\label{sec:KSVD}

The proposed model extracts words, n-grams, skip-grams or bags-of-words of each tweet and uses them as features to determine whether an event will occur on a specific day. The features are counted on a daily basis into vectors representing the times series of the keyword volume. The feature-vectors are then correlated with the vector of daily events. 

The total number of words extracted as features exceeds 10 million per day, rendering most data processing techniques computationally infeasible. To solve this issue, the features are filtered to exclude those with very low correlations, which comprise the majority of the features processed. We retain only the 10,000 features with the highest correlations for further processing. The challenge in dealing with the remaining data is that individual features have relatively low correlation scores, which implies a low association between features and events.

Enhancing the correlation between the textual features and the events requires finding new features with time-series highly associated with the event-time-series. To accomplish this, we propose grouping semantically similar features into a combined feature, and calculate the sum of the similar features' vectors into a single vector representing them all. Here we use the idea of matrix factorization that is used in LSI, but for the purpose of clustering rather than finding the most similar documents (or days in this case). We analyze the relation between events (i.e. protests) and the features used (words, BOWs, etc) on each day to produce a set of concepts relating the features used to the events of the day. 

\begin{figure*}
% Use the relevant command to insert your Figure file.
% For example, with the graphics package use
\centering
  \includegraphics[width=0.80\textwidth]{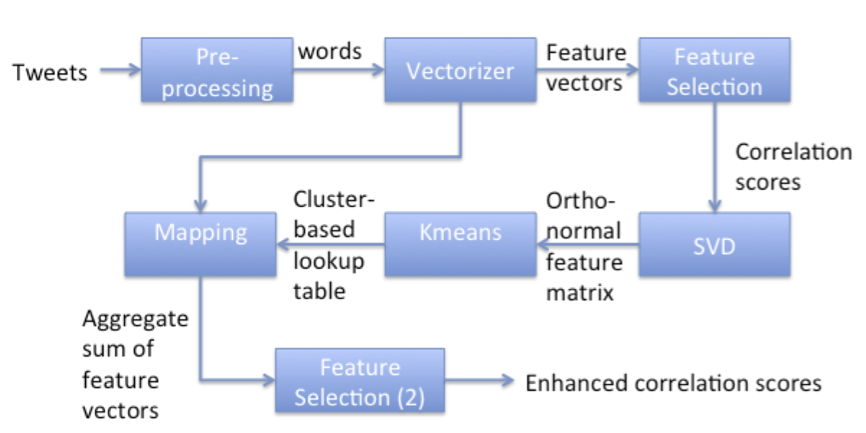}
% Figure caption is below the Figure
\caption{The pipeline to extract the textual features, correlate them with events then use SVD and K-means to enhance the correlation by merging the related weak features.      }
\label{fig:pipeline}       % Give a unique label
\end{figure*}

We assume that words with similar meaning are more likely to occur in similar contexts (i.e. days) with similar variability according to the distributional hypothesis. So, we formulate the feature-count-per-day relation as a matrix, where features are represented as rows and days are represented as columns. Then, we use singular value decomposition (SVD) to decompose the matrix into a features matrix, a day matrix and a singular matrix mapping the two matrices to each other. After decomposition, instead of measuring the distance with the daily vectors to cope with LSI, we will cluster the feature matrix to create a look-up table mapping the features within each cluster to its centroid.

The proposed technique consists of five steps to be applied after the initial selection of features. The first step is ensuring feature independence using singular value decomposition (SVD). SVD factorizes the feature/day matrix into a feature matrix ($U$) and observation matrix ($V$) and a singular matrix mapping the features to observations according to the equation \ref{eq:svd1} 

\begin{equation}
\label{eq:svd1}
X= U\Sigma V^T
\end{equation}

\noindent where, 
\begin{itemize}
\item     $m$ is the number of features and $n$ is the number of days.
\item     $A$ is an $m \times n$ matrix representing the feature/day vectors.
\item     $U$ is an $m \times m$ matrix representing the feature vectors. This matrix is unitary and orthogonal.
\item     $\Sigma$ is a diagonal m × n matrix of non-negative real numbers.
\item     $V^T$ is the transpose of the $n \times n$ unitary matrix ($V$), representing the days.
\end{itemize}

The values of the diagonal matrix $\Sigma$ are the singular values of the original matrix $A$. These singular $\sigma_i$ values are usually listed in a descending order. The singular values determine the strength of its related vector in $U$ as a feature, as formulated in equation \ref{eq:svd2}.
\begin{equation}
\label{eq:svd2}
\begin{bmatrix}
x_{1,1} & \dots & x_{1,n} \\ \vdots & \ddots & \vdots \\ x_{m,1} & \dots & x_{m,n}
\end{bmatrix}
=
\begin{bmatrix}
\begin{bmatrix} ~\\u_1\\~\end{bmatrix}& \dots & \begin{bmatrix} ~\\u_m\\~\end{bmatrix}\\ 
\end{bmatrix} 
\begin{bmatrix}
\sigma_{1,1} & \dots & 0 \\ \vdots & \ddots & \vdots \\ 0 & \dots & \sigma_{m,n}
\end{bmatrix}
\begin{bmatrix}
\begin{bmatrix} ~~~& v_1 &~~~\end{bmatrix}\\   \vdots  \\ \begin{bmatrix} ~~~& v_n &~~~\end{bmatrix}
\end{bmatrix}
\end{equation}

Since $U$ and $V^T$ are unitary, the columns of each of them form a set of orthogonal vectors, which can be considered as basis dimensions. The matrix $\Sigma$ maps the basis dimensions of $v_i$ to the vector $u_i$ after being stretched using $\sigma_i$. Since $U$, $U^T$, $V$ are unitary matrices and their columns are orthogonal, we can measure the distance between any two features using Euclidean distance. This measures the similarity between any two words considering the context of the original matrix.

The second step is to cluster the independent features using k-means in order to partition the orthogonalized features into a set of clusters $S = \{S_1,S_2,S_3,\dots ,S_k \},$ with size $k$. The objective is to minimize the pairwise distance of points within the same cluster, by minimizing the sum of squares in each cluster. The objective function is formulated by equation \ref{eq:kmeans}: 

\begin{equation}
\label{eq:kmeans}
\argmin_S \sum_{i=1}^{k} \sum_{u \in S} {||u-\mu||}^2 = \argmin_S \sum_{i=1}^{k} |S_i| Var S_i
\end{equation}

\noindent where $S$ is the set of $k$ clusters and $\mu_i$ is the mean of points in $S_i$.
The clusters are initialized using multiple random partitioning and the distance between any two points is calculated using the Euclidean distance. The Lloyd algorithm is used for $k$-means, consisting of the two steps for assignment and update as described below.

Assignment step: To assign each data item to the cluster that has the closest mean value.

\begin{equation}
\label{eq:LloydStep1}
S_i^{(t)}=\{\, u_p~ : ~ {||u_p-m_i^{(t)}|| }^2<{||u_p-m_j^{(t)} ||}^2 ~~ \forall ~~ j, 1<j<k \,\}
\end{equation}

Update step: To find the new cluster centroid that achieves the minimum distance with all other data items within the cluster.

\begin{equation}
\label{eq:LloydStep2}
m_i^{(t+1)}=  \dfrac{1}{|S_i^{(t)}|} \sum_{u_j\in S_i^{(t)}}u_j 
\end{equation}

Although the algorithm achieves relatively good clustering results,  it does not guarantee to achieve the optimum solution, as it is an NP-Hard problem.

The clusters resulting from the $k$-means are used to build a look-up table mapping the features in each cluster to the cluster centroid that represents the contents of the whole cluster. The third step is to apply this mapping to the original data, where all the signals of all the words in each cluster are summed to the cluster centroid using equation \ref{eq:merge}. 
\begin{equation}
\label{eq:merge}
c_i= \sum_{x_j \in C_i} x_{j,1...n}     
\end{equation}

\noindent where  $i \in \{1,\dots k\}$  and $C_i$ is the set of the original raw non-orthogonal vectors associated with the keywords that belong to the cluster $S_i$ resulting from $k$-means, and $c_i$ is the sum of all vectors in $C_i$.

The last step is to recalculate the correlation scores after the aforementioned summation. This process increases the correlation scores for the cluster centroids, which promises better results for classification or prediction purposes. 

\section{Experiment and Results}
\label{subsec:Dataprep}

The experiments are designed to calculate the correlations between the term frequency vector of the features and the frequency vector of the civil unrest events within a specific time-frame. In our experiments, we will consider words of different forms and counts as our features and the count of the civil unrest events as our golden standard record (GSR). Afterwards, the proposed technique is applied by decomposing the feature/day matrix $X$ to extract the feature matrix $U$. The feature matrix $U$ is clustered using k-means to build the lookup table. The look-up table is used to merge the features within each cluster by adding the sum of the vectors of all features within a cluster to the vector of the centroid BOW of the cluster. 

The data used in this experiment consist of the tweets used as a predictor for the future events and the news used as a descriptor for the events already happened along the same timeframe. The tweets are collected from Twitter using the GNIP service where we bought all the tweets issued by any user within Australia for the studied time frame. These tweets are furtherly processed to extract the most informative features that can be used to classify the day as event/non-event days. The news articles reporting the events occurred are automatically collected using RSS feeds and manually labelled using a set of field experts from police and intelligence that identified the civil unrest events of interest. This news are used as our golden truth to compare our classification results with. The collected data can be described as follows:

\begin{itemize}
\item The time frame is 640 days of tweets that are mapped to 640 days of news articles reporting civil unrest events
\item  Each day has 3 million tweets on average in Melbourne, 3.5 million tweets in Sydney, 2 million tweets in Brisbane, 1 million tweets in Perth and 500 thousands tweets in Adelaide.
\item  Each tweet has 10 words on average, which forms 90 BOWs per tweet, which form 270 million BOWs per day in Melbourne
\item  Aggregating similar BOWs by summing the counts of similar bows will reduce the total number of BOWs to less than 50%  
\item  Filtering out all Bows with small counts can eliminate more than 90\% of the BOWs according to the filter threshold, in our experiment we eliminate any bow occurred for less than 5 times per day. 
\item  The resulting total number of BOWs is to be used as features is around 10 million BOWs per day.
\item  The total number of BOWs used along the whole time frame is 6400 million BOW. 
\end{itemize}

The experiment has been performed on a time frame of 640 days within the geographical area of Melbourne. The location of the tweets is determined using (1) tweet location, (2) the longitude and latitude, (3) the time zone and (4) the profile location. The first step is preprocessing, where we clean and prepare the data and extract the BOWs for correlation. Data preparation is a multi-step process that includes data cleaning, NLP analysis, word counts and GSR preparation. Example 1 shows how the tweet is cleaned, prepared and vectorized to be used in correlation. These steps are explained as follows: 

%\subsection{Data preparation}
%\label{subsec:Dataprep}

\begin{enumerate}
\item The data is cleaned by excluding all tweets in any language other than English and all tweets with any URLs; and removing non-Latin characters, hashtags, HTML tags, punctuation, and stopping words (using the NLTK list \cite{Loper:2002NLTK}) from the remainder.
\item Each tweet is split into a list of features with different lengths varying between one and three as follows:
    \begin{enumerate}
      \item{Keywords:} Each individual word within the tweet .
      \item{N-Grams}: Any N contiguous words in the tweet.
      \item{Skip-Grams}: Any N non-contiguous words within the same tweet  in the same order (e.g. [``march'', ``melbourne''] is a different feature than [``melbourne'',``march'']).
      \item{Bags-of-Words}: Any non-contiguous N words within the same tweet irrespective of order (e.g. [``march'', ``melbourne''] is exactly the same feature as [``melbourne'',``march'']).
    \end{enumerate}

\item All words in each BOW are lemmatized using the NLTK lemmatizer in order to return each word to its origin, to avoid grammatical effects on the word shape (e.g. "Went" $\rightarrow$ "Go").
\item After lemmatization, all words in each BOW are stemmed using the Lancaster stemmer in order to return similar words to their dictionary origin (e.g. "Australian" $\rightarrow$ "Austral").
\item Each BOW is counted in the tweets of Melbourne for each day to prepare the term frequency vectors.
\item Load the press events as GSR and count them per day for the whole time-frame.
\end{enumerate}

\lstset{
  basicstyle=\ttfamily,
  columns=fullflexible,
  keepspaces=true,
}
\begin{lstlisting}[frame=single,mathescape,breaklines=true]
Example 1: 
Tweet: Highlight sign from #KeepSydneyOpen march:"My Friends Have Gone To Melbourne"
Tweet after cleaning and stemming: highlight sign march friend go melbourn

List of BOWs: [highlight, sign], [highlight, march], [highlight, all],  [highlight, friend], [highlight, go], [highlight, melbourn],[sign, march], [sign, all],  [sign, friend], [sign, go], [sign, melbourn],  [march, all],  [march, friend], [march, go], [march, melbourn],  [all, friend], [all, go], [all, melbourn], [friend, go], [friend, melbourn], [go, melbourn]

[highlight, sign]: [$x_{1,1}$,$x_{1,2}$,$x_{1,3}$,$\dots$,$x_{1,640}$]
[highlight, march]:  [$x_{2,1}$,$x_{2,2}$,$x_{2,3}$,$\dots$,$x_{2,640}$]
[highlight, all]:  [$x_{3,1}$,$x_{3,2}$,$x_{3,3}$,$\dots$,$x_{3,640}$]
.
.
.
[sign, march]: [$x_{4,1}$,$x_{4,2}$,$x_{4,3}$,$\dots$,$x_{4,640}$]
[sign, all]: [$x_{5,1}$,$x_{5,2}$,$x_{5,3}$,$\dots$,$x_{5,640}$]
[sign, friend]: [$x_{6,1}$,$x_{6,2}$,$x_{6,3}$,$\dots$,$x_{6,640}$]
[sign, go]: [$x_{7,1}$,$x_{7,2}$,$x_{7,3}$,$\dots$,$x_{7,640}$]
GSR (Event Count per day) : [$e_{1}$,$e_{2}$,$e_3$,$\dots$,$e_{640}$]
\end{lstlisting}

%\subsection{Initial correlation}
%\subsection{Correlate, Decompose, cluster, map}
%\label{subsec:Correlation}
The second step is to correlate the vectors of each word with the GSR vector and select the 10,000 words with highest correlation. The correlation process is described in equation \ref{eq:correlation1}, where $v$ is the term frequency for a specific word in a specific feature, $e$ is the number of civil unrest events that took place in a single day, and $c$ are the correlation scores between each word's frequency row and the vector of the event count. The top 10,000 words are selected to reduce the computational complexity of the matrix decomposition in the next step.
\begin{equation}
\label{eq:correlation1}
\begin{bmatrix}
x_{1,1} & \dots & x_{1,n} \\ \vdots & \ddots & \vdots \\ x_{m,1} & \dots & x_{m,n}
\end{bmatrix}
\otimes
\begin{bmatrix}
e_1 \\ \vdots \\e_n
\end{bmatrix}
=
\begin{bmatrix}
c_1 \\ \vdots \\c_n
\end{bmatrix}
\end{equation}

\noindent where $x$ is the feature count per day. Rows represent the features and columns represent the day. $e$ is the number of civil unrest events that happened in each day, $c$ is the correlation between each word and event vector, and $\otimes$ is the correlation method used in each experiment, such as the Pearson or Spearman correlation.
%\subsection{Decomposition and Clustering}
%\label{subsec:Enhancement}

\begin{table}[]
\centering
\caption{The correlation scores of the top correlated word before and after applying the proposed technique for each of the five correlation techniques. }
\label{tbl:max_scores}
\begin{tabular}{|l|r|r|r|r|r|r|r|r|r|r|}
\hline
            & \multicolumn{2}{c|}{Pearson}                             & \multicolumn{2}{c|}{Spearman}                            & \multicolumn{2}{c|}{Kendal Tao}                          & \multicolumn{2}{c|}{Distance correlation}                & \multicolumn{2}{c|}{Mutual Info}                         \\ \hline
            & \multicolumn{1}{c|}{before} & \multicolumn{1}{c|}{after} & \multicolumn{1}{c|}{before} & \multicolumn{1}{c|}{after} & \multicolumn{1}{c|}{before} & \multicolumn{1}{c|}{after} & \multicolumn{1}{c|}{before} & \multicolumn{1}{c|}{after} & \multicolumn{1}{c|}{before} & \multicolumn{1}{c|}{after} \\ \hline
UniGram     & 0.302                       & 0.751                      & 0.215                       & 0.425                      & 0.205                       & 0.376                      & 0.264                       & 0.808                      & 0.811                       & 0.863                      \\ \hline
Bi-gram     & 0.314                       & 0.714                      & 0.241                       & 0.464                      & 0.231                       & 0.425                      & 0.291                       & 0.723                      & 0.533                       & 0.827                      \\ \hline
Tri-gram    & 0.284                       & 0.648                      & 0.214                       & 0.381                      & 0.205                       & 0.346                      & 0.252                       & 0.691                      & 0.512                       & 0.723                      \\ \hline
Skip-gram-2 & 0.308                       & 0.645                      & 0.241                       & 0.650                      & 0.231                       & 0.542                      & 0.286                       & 0.744                      & 0.705                       & 0.854                      \\ \hline
Skip-gram-3 & 0.308                       & 0.632                      & 0.224                       & 0.580                      & 0.214                       & 0.541                      & 0.258                       & 0.707                      & 0.551                       & 0.873                      \\ \hline
BOW-2       & 0.310                       & 0.669                      & 0.260                       & 0.621                      & 0.244                       & 0.525                      & 0.299                       & 0.759                      & 0.702                       & 0.851                      \\ \hline
BOW-3       & 0.327                       & 0.515                      & 0.274                       & 0.720                      & 0.261                       & 0.683                      & 0.284                       & 0.814                      & 0.699                       & 0.850                      \\ \hline
\end{tabular}
\end{table}

\begin{table}[]
\centering
\caption{The average correlation scores of the top 100 correlated words before and after applying the proposed technique for each of the five correlation techniques.}
\label{tbl:avg_scores}
\begin{tabular}{|l|r|r|r|r|r|r|r|r|r|r|}
\hline
            & \multicolumn{2}{c|}{Pearson}                             & \multicolumn{2}{c|}{Spearman}                            & \multicolumn{2}{c|}{Kendal Tao}                          & \multicolumn{2}{c|}{Distance correlation}                & \multicolumn{2}{c|}{Mutual Info}                         \\ \hline
            & \multicolumn{1}{c|}{before} & \multicolumn{1}{c|}{after} & \multicolumn{1}{c|}{before} & \multicolumn{1}{c|}{after} & \multicolumn{1}{c|}{before} & \multicolumn{1}{c|}{after} & \multicolumn{1}{c|}{before} & \multicolumn{1}{c|}{after} & \multicolumn{1}{c|}{before} & \multicolumn{1}{c|}{after} \\ \hline
UniGram     & 0.228                       & 0.221                      & 0.182                       & 0.207                      & 0.173                       & 0.189                      & 0.1993                      & 0.213                      & 0.687                       & 0.732                      \\ \hline
Bi-gram     & 0.244                       & 0.285                      & 0.183                       & 0.310                      & 0.175                       & 0.292                      & 0.205                       & 0.383                      & 0.314                       & 0.420                      \\ \hline
Tri-gram    & 0.230                       & 0.188                      & 0.172                       & 0.246                      & 0.164                       & 0.221                      & 0.197                       & 0.269                      & 0.178                       & 0.249                      \\ \hline
Skip-gram-2 & 0.269                       & 0.433                      & 0.206                       & 0.437                      & 0.197                       & 0.411                      & 0.228                       & 0.537                      & 0.445                       & 0.620                      \\ \hline
Skip-gram-3 & 0.268                       & 0.374                      & 0.200                       & 0.439                      & 0.192                       & 0.411                      & 0.223                       & 0.425                      & 0.496                       & 0.293                      \\ \hline
BOW-2       & 0.270                       & 0.434                      & 0.210                       & 0.414                      & 0.200                       & 0.394                      & 0.231                       & 0.520                      & 0.472                       & 0.659                      \\ \hline
BOW-3       & 0.286                       & 0.457                      & 0.219                       & 0.593                      & 0.209                       & 0.563                      & 0.247                       & 0.590                      & 0.490                       & 0.643                      \\ \hline
\end{tabular}
\end{table}

After selecting the top correlated words from the correlation step, we will decompose the matrix of the selected words/days using SVD and use the feature representation matrix $U$ from equation \ref{eq:svd1} for clustering in the next step. Although SVD is usually used for feature reduction we will use all the features in the clustering step to build a mapping table. Then we apply the k-means to find 1000 clusters, which gathers each word with 10 other words. A smaller cluster makes the resulting signal weaker, while a larger cluster makes the signal noisy as it will include unrelated component signals that will corrupt each other. %The result of the clustering process is a table where each centroid word represents multiple other related words as shown in table \ref{tbl:mapping}.

The clustering process set the target number of clusters ($k$) to $2000$ in order to have 5 words per cluster on average, though the cluster size is not guaranteed in k-means. The centroids are seeded to the algorithm using random numbers for 50 runs. The maximum number of iterations per each run is set to 35 as most of the runs saturate before the 25th iteration. This technique increased the maximum correlation scores for the selected centroids of the 2000 clusters from an average of 0.3 to 0.65 for Pearson correlation, from 0.23 to 0.54 for Spearman correlation, from 0.22 to 0.49 for Kendal Tao correlation, from 0.27 to 0.74 for distance correlation and from 0.64 to 0.83 for mutual info. The maximum correlation scores for each combination of the correlation-method, word-form and word-count before and after applying the proposed method are stated in table \ref{tbl:max_scores}. The average correlation scores for the same combinations before and after applying the method are stated in table \ref{tbl:avg_scores}. 

The same experiment is applied to five cities in Australia, and the correlations are enhanced with various margins, where Sydney achieved the highest marginal  in the correlation scores and Adelaide achieved the least enhancement in the correlation scores, we also applied the same experiment using the Indonesian language in the city of Jakarta and the correlation scores are enhanced as well, even though we did not use any Indonesian lemmatizer or stemmer. The top correlated BOWs for the civil unrest are listed below, where the words are lemmatized and stemmed using Lancaster stemmer. Most of the top correlated BOWs are related to protests in Melbourne, or related to some political figure or some cause.  But, some other BOWs are not related to civil unrest as they got accidental spurious correlation with the events days because of some statistical bias. For example, Most protesters prefer to act on Mondays, while some TV show is displayed each Monday, this will lead to spurious correlation and fake association. 

\begin{lstlisting}[frame=single,mathescape,breaklines=true]
List of top 25 BOWS associated with the civil unrest events in Melbourne 
['newsmelb', 'protest'], ['ral', 'tram'], ['park', 'permit'], ['middl', 'typ'], ['fuck', 'jfc'], ['connect', 'wom'], ['bib', 'kam'], ['caulfield', 'look'], ['ral', 'rout'], ['pettifl', 'son'], ['keynot', 'lik'], ['awkward', 'quest'], ['forget', 'land'], ['afl', 'strong'], ['limit', 'step'], ['hono', 'lif'], ['newsmelbourn', 'protest'], ['felt', 'weekend'], ['detail', 'seem'], ['joel', 'say'], ['afraid', 'ground'], ['nya', 'resel'], ['revolv', 'watch'], ['navbl', 'rwy'], ['energy', 'kil']
\end{lstlisting}

\section{Discussion}

The correlation scores for single words as features were too low and were not expressive. The single word data is noisy and misused frequently within the different contexts. Additionally, many words had relatively high correlations due to coincidence. 

The n-grams produced slightly higher correlations without effective significance. The n-gram vectors had a high number of zeros because of the low probability of the same word-sequence being repeated multiple times with the same pattern. The higher number of n-grams led to a lower probability of re-occurrence, higher frequency of zeros and lower counts per vector. The signal was too weak to use the n-gram as a feature.

The skip-gram produced slightly higher correlations than keywords and n-grams. The main advantage is to maintain the context of the word by pairing it with its co-occurring words in the same context. The number of zeros was lower than that of the n-gram method and higher than the keyword method. The counts per day were slightly higher as well, which strengthened the signal without increasing noise. The best correlation scores are achieved for 2-word skip-grams due to the highest counts (resulting in the strongest signal). Increasing the number of words per skip-gram leads to a higher number of zeros and lower counts, which weakens the signal significantly.

Bags-of-words are the best content–based feature so far as they produce the highest correlation scores as well as the highest number of correlated BOWs. The bags-of-words have limited zeros and high counts, which imply a signal stronger than the other word forms. BOWs also consider the words' co-occurrences per tweet, which preserves the contextual meaning of each word. The size of the BOW affects the strength signal as well, where 2-words BOWs achieved higher correlation, stronger signal and lower noise.

3-word and 4-word bags of words are too limited in their data set to be used as predictors or to be correlated with the GSR events. The vector of counts for each BOW has too many zeros and small values. It is highly improbable to have the same exact four words in multiple tweets unless considering retweets and embedded tweets. This causes the daily counts of a 4-word BOW to be low enough that accidental text and spurious data are significant factors. Filtering the data to avoid spurious BOWs leads to filtering most other BOWs as well, causing aggressive limitation of the number of potential predictors. Although it is recommended to use 5-word BOWs in topic modelling problems, it is not practical for Twitter due to the limited number of characters per tweet, as 140 characters make around 16 words after excluding URLs, hash-tags and mentions.

\section{Conclusion}
In this paper, we proposed to enhance the correlation of textual features gathered from Twitter with civil unrest events by combining related features into one. This combination is implemented by finding the sum of the vectors of the related features element-wise. In order to identify which features are related to each other, we proposed to decompose the feature/event matrix using SVD then cluster the feature matrix using k-means. The importance of SVD is that it guarantees the features' locations in the space are mapped to orthogonal dimensions, which isn't always the case in the original matrix as neither the features nor the days are guaranteed to form orthogonal dimensions. As k-means uses Euclidean distance and cannot work without orthogonal dimensions, this is necessary. Each cluster will be represented using one feature that has the minimum distance to all other features within the same cluster. The cluster is used to build a look-up table mapping each feature to the centroid feature of its cluster. This look-up table will be used to determine which vectors to sum together using the raw (not-decomposed) vectors.

The experiments and results showed that the proposed technique increased the correlation scores for the centroid of the clusters significantly, with an average increase in correlation score of 0.3. This technique has been tested for multiple correlation techniques including Pearson, Spearman, Kendal Tao, distance correlation and mutual information, and increased correlation scores for all five.

The future work is to try a clustering technique that guarantees equally sized clusters, and to try to eliminate any unrelated keywords within the cluster that may have appeared because of the spurious nature of the data. This method can also be tested on other feature selection and data association techniques such as the maximal information coefficient, cosine similarity index and Jaccard similarity index.

\bibliographystyle{spmpsci}      % mathematics and physical sciences
\bibliography{references}   % name your BibTeX data base

\end{document}